\documentclass[doublespacing]{elsart}
\usepackage{natbib}
\usepackage{epsfig}

\newcommand{\be}{\begin{equation}}
\newcommand{\ee}{\end{equation}}

\begin{document}
\begin{frontmatter}
\title{Caloric Curves in two and three-dimensional Lennard-Jones-like systems 
including Long-range forces}
\author[df]{M.J.Ison\corauthref{cor1}}
\ead{mison@df.uba.ar}
\author[df]{A.Chernomoretz}
\author[df]{C.O.Dorso}
\corauth[cor1]{Corresponding author.}

\address[df]{Departamento de F\'{\i}sica-Facultad de Ciencias Exactas y
Naturales, Universidad de Buenos Aires, Pabell\'on 1 Ciudad Universitaria,
1428 Buenos Aires, Argentina}
\thanks[]{Partially supported by the University of Buenos Aires via grant X139}

\begin{abstract}
We present a systematic study of the thermodynamics of two and three-dimensional
generalized Lennard-Jones ($LJ$) systems focusing on the relationship
between the range of the potential, the system density and its dimension.
We found that the existence of negative specific heats depends on these
three factors and not only on the potential range and the density of the
system as stated in recent contributions. 
\end{abstract}

\end{frontmatter}

\section{Introduction}

In the last few years a lot of attention has been paid to Hamiltonian
systems in which the interaction potential range is of the order of the size
of the system. Typical examples include nuclei, metallic clusters and
galaxies. In the first two cases the system under consideration is small
comprising just a few (or less) hundreds of particles, while in the third
one the range of the potential is long \cite{tsa1,curilef,tsapre,tamarit}.

One of the main consequences of dealing with long-range Hamiltonian
systems is the appearance of negative specific heats. Negative specific
heats appear in the literature of small systems both in experimental \cite
{D'agostino,Haberland} and theoretical \cite{Labastie,gross} studies. In
particular, in a series of recent papers, we have found such a behavior for
highly excited small Lennard Jones aggregates ($147$ particles) both free to
expand and constrained in an spherical volume \cite{noneqfrag,mison}. In
this case the standard ($12-6$) $LJ$ systems in $3$ dimensions qualifies as
a short-ranged potential, since the exponent of the attractive term of the 
potential ($\alpha=6$) is greater than the dimension of the system ($d=3$). 
On the other hand, in \cite{tsa1} it is claimed, based on the numerical 
analysis of two-dimensional systems, that a negative specific heat region is 
present only if long-range forces are present. It is therefore necessary to 
make a complete study in order to clarify the interplay between the potential 
range and the thermodynamics of the system.

The presence of such a long range interaction in the system calls for a
proper thermodynamical description. One can, for instance, try to use
Tsallis statistics \cite{tsallis}. However, ''small'' systems can be also
studied via micro-canonical statistics \cite{gross}. Our studies and
conclusions will be referred to the $N,V,E$ ensemble.

This work is organized as follows. In Section 2 we present the model, the
interaction potential and the observables that we use in the calculations.
In Section 3 we show the results of our simulations for two and 
three-dimensional systems. Finally in Section 4 we present our conclusions.

\section{The model}

\begin{figure}[htbp]
\centerline{%
\hbox{       
   \psfig{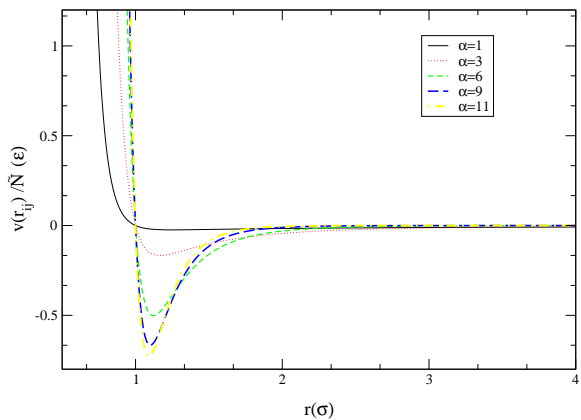}}}
\caption{\textit{Set of generalized $LJ$ $3D$ potentials}.}
\label{Vvsr}
\end{figure}

The system under study is composed by a gas of $147$ particles confined in a
spherical box, defined by the Hamiltonian $H=K+V/\tilde{N}+V_{walls}$, where
K is the kinetic energy and the interaction potential is given by $V=\Sigma
v(r_{ij})$ with $v(r_{ij})=C_\alpha [ \frac{\sigma}{r_{ij}}^{-12}-\frac{%
\sigma}{r_{ij}}^{-\alpha}] (0 \leq \alpha < 12)$ and $C_\alpha= \epsilon
(12^{12}/\alpha^{\alpha})^{1/(12-\alpha)}/(12-\alpha).$ The Lennard-Jones
like potentials as a function of $r$ are shown in Fig. \ref{Vvsr} for $d=3$, 
$N=147$, and several values of $\alpha$. We used a spherical confining wall
via an external potential $V_{wall}\sim (r-r_{wall})^{-12}$ with a cut off
distance $r_{cut}=1\sigma $. Energies are measured in units of the potential
well ($\epsilon $), and the distance at which the potential changes sign ($%
\sigma$), respectively. The unit of time used is $t_{0}=\sqrt{%
\sigma^{2}m/48\epsilon }$.

The nonextensive scaling parameter, introduced in \cite{tsa1}, $\tilde{N}%
\equiv 1+d\int_{1}^{N^{1/d}}drr^{d-1-\alpha }$ is convenient to make the
Hamiltonian \textit{formally} extensive $\forall \alpha /d$ ($d$ is the
dimensionality of the system).

\smallskip

The model is exactly the one introduced in \cite{tsa1}, and the particular
case $\alpha =6$ recovers the celebrated Lennard-Jones model.

\smallskip

The set of classical equations of motion were integrated using the velocity
Verlet algorithm, which preserves volume in phase space \cite{allen}; with an
integration time step between $0.001t_{0}$ and $0.01t_{0}$ which guaranteed
a conservation of the energy not worse than $0.01\%$. Initial conditions
were constructed from the ground states of $LJ$ systems in $r-space$ and
rescaling velocities with a Maxwellian distribution of velocities to the
desired value of energy. All calculations were performed once the transient
behavior was over.

\subsection{Discussion about the $\tilde{N}$ term}

To properly understand the inclusion of the $\tilde{N}$ normalization term
let's remember the cluster expansion in real gases. By integrating in
moments we end up with the proportionality of the partition function and the
configuration integral 
\begin{equation}
\frac{1}{\lambda ^{3N}N!}\int {d^{3N}qexp[-\beta \Sigma v_{ij}]}
\end{equation}
In the Van der Waals approximation 
\begin{eqnarray}
V(r) &=&\infty \quad \quad \quad r<r_{0} \\
exp[-\beta V(r)] &\sim &1-\beta V(r)\quad r>r_{0}
\end{eqnarray}

It can be easily shown that 
\begin{equation}
E-E_{ideal}=N_{pairs} <V(r)>
\end{equation}

with $<V(r)>=\frac{4\pi}{V} \int{dr r^2 V(r)}$. We see that $\tilde{N}$ is a
generalization of this term taking into account the adimensionalization and
the finite number of particles. As the configurational integral is
ill-defined for long-range potentials (decaying slower than $d$)
thermodynamic quantities $A_N$ like the internal energy, the free energy,
Gibbs energy, etc. associated with systems including long-range potentials
scale like \cite{curilef} 
\begin{equation}
\frac{A_N}{N} \propto \int_1^{N^{1/d}} dr r^{d-1-\alpha}=\frac{1}{d} \\
\frac{N^{1-\alpha/d}-1}{1-\alpha/d}
\end{equation}

By normalizing the potential energy with $\tilde{N}$ the Hamiltonian is
turned formally extensive.

\subsection{The observables}

The main observable to be extracted from our simulations is the caloric
curve ($CC$), which is defined as the functional relationship between the
temperature of the system and its energy in terms of the density i.e. $%
T(E,\rho )$, from which we define the specific heat as

\begin{equation}
C_{v}=\frac{1}{\partial T/\partial E)_{v}}
\end{equation}

It is then clear that a $C_{v}<0$ will be obtained if and only if the $CC$
displays a loop.

It has recently been proposed that first order phase transitions would be
univocally signed by the amount of fluctuations in the different subsystems
in which the system can be subdivided \cite{gulmi_fluc}. In particular, the
relative kinetic energy fluctuation $A_{K}$ is defined as:

\begin{equation}
A_K=\frac{N \sigma^2_K}{T^2}  \label{eqA}
\end{equation}

where $N$ is the number of particles, $\sigma_K$ the standard deviation of
the kinetic energy per particle and $T$ the temperature of the system. Since
kinetic energy fluctuations and the specific heat are related by \cite
{lebowitz}

\begin{equation}
N <\sigma_K^2>_E={\frac{d}{{2 \beta^2}}} (1 - {\frac{d }{{2 C}}})
\label{eqFluc}
\end{equation}

Negative values of the specific heat should be expected whenever $A_K$
exceeds the canonical value ($A_K=1.5 \epsilon$ for $d=3$ and $A_K=1.0
\epsilon$ for $d=2$) \cite{gulmi_fluc, noneqfrag}.

\section{Results}

\subsection{Caloric Curves and Kinetic Energy Fluctuations in three
dimensions}

\begin{figure}[htbp]
\centerline{%
\hbox{       
   \psfig{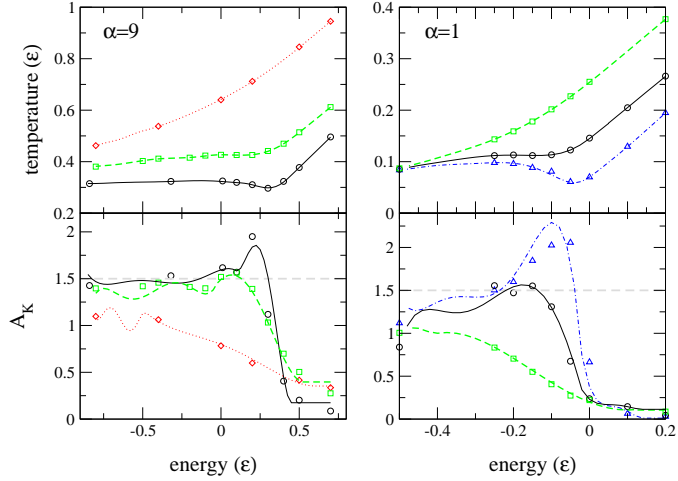}}}
\caption{\textit{Caloric curves and relative kinetic energy fluctuations for 
$\alpha=9$, $\rho=0.01 \sigma^{-3}$ (circles), $\rho=0.06 \sigma^{-3}$
(squares) and $\rho=0.3 \sigma^{-3}$ (diamonds) (left panels) and $\alpha=1$%
, $\rho=0.001 \sigma^{-3}$ (triangles), $\rho=0.01 \sigma^{-3}$ (circles), $%
\rho=0.06 \sigma^{-3}$ (squares) (right panels).} }
\label{ccA}
\end{figure}

We will first focalize in the study of three-dimensional systems. In Fig.\ref
{ccA} we show the caloric curve for a system composed of $147$ particles
with $\alpha =9$ (upper left panel) and for $\alpha =1$, a long-range
potential (upper right panel). For future reference, it is interesting to
notice that the case $\alpha =9$ for 3 dimensions would correspond to $%
\alpha =6$ in 2 dimension because for both the quotient $\frac{\alpha }{d}%
=3. $ It can be seen that a loop in the caloric curve is present in both
cases only when the system density is low (notice the density differences
in both panels). For a short-ranged potential the relationship between the
caloric curves and the constraining volume has already been clarified (see 
\cite{noneqfrag}): The presence of a loop in the caloric curve can be
related to the formation of drops in configuration space. When dealing with 
long-ranged potentials the scenario slightly changes, even though the clusters 
can not be completely isolated, we still have weakly-interacting clusters, 
they just need more room to accommodate. We will come back to this point later.

\begin{figure}[htbp]
\centerline{%
\hbox{       
   \psfig{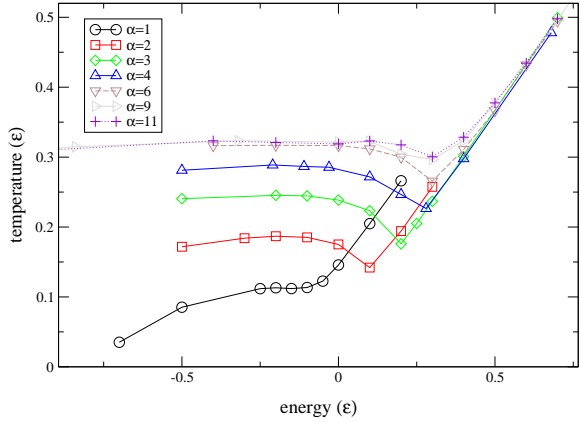}}}
\caption{\textit{Range dependence of the caloric curves for $\rho=0.01
\sigma^{-3}$, $d=3$}. }
\label{cc3d}
\end{figure}

In the lower panels of Fig. \ref{ccA} we show $A_{K}$ for several densities
(see caption for details). Whereas the symbols refer to calculations of $A_K$
via fluctuations (Eq.\ref{eqA}) the lines show the corresponding $A_K$ 
estimations using the information provided by the caloric curves, through 
Eq.\ref{eqFluc}. It is immediate that fluctuations are enhanced well above the 
canonical value for the cases where the caloric curve displays a loop.

In Fig.\ref{cc3d} we show the dependence of the caloric curve with the range
of the interaction for a constant density ($\rho =0.01\sigma ^{-3})$, which
correspond to a low density for all but the extremely long-range potential $%
(\alpha =1)$, reflected by the absence of a loop in the caloric corresponding
curve. Also notice that the energy which points the entrance of the system into 
the vapor branch is an increasing function of the total energy for long-range 
systems ($\alpha \leq 3$) and then collapses to a constant value for the 
short-range cases studied.

\subsection{Statistical inefficiency corrections}

The importance of the loop in the caloric curve resides in the fact that it
is a signal which identifies the presence of a first order phase transition
for the corresponding density and lower ones. In Ref \cite{cheba} the phase
diagram for a pure $LJ$ system was constructed by analyzing different
observables. As a consequence of this, the phase diagram was divided into 
three density regions. In the low density regime the $CC$ displays a loop 
whereas for higher densities the identification of the transition line was 
performed by a phase-space analysis. It is therefore relevant to study the 
relationship between the range of the potential and the density for which the 
loop disappears which we have called 'disjoint density $\rho _{dis}$' since it
acts like a separatrix between two regions of the phase diagram.

Up to this point we have identified the presence of negative specific heats
via two independent signals: The presence of the loop in the $CC$ and
abnormally large kinetic energy fluctuations $A_{K}$. We define $\rho _{dis}$
as the density at which the signals (presence of the loop including errors
and relative fluctuations greater than the canonical value) either both
disappear at the same time or they become inconsistent, i.e. a loop without
big relative fluctuations or viceversa. It is worth to notice that only in a
region in $\rho $-space very close to $\rho _{dis}$ these two signals
can become inconsistent.

Special care should be taken because not only we are dealing with
observables linked to mean values but to fluctuations (second moments), as
well. In order to disregard erroneous results we performed a statistical
inefficiency study, a tool which was introduced by Jacucci and Rahman \cite
{rahman} to estimate errors from correlated data series.

The block average procedure is used to calculate the error of a quantity $K$
computed from a correlated data series. In order to establish the error, the
data are grouped into $b$ blocks of length $n_b$. We expect the averages of
the data in each block no to be correlated for sufficiently long $n_b$. The
block size ($n_b$) is used to obtain uncorrelated data (the statistical
inefficiency $s$) for a quantity $K$ using:

\begin{equation}
s = \lim_{n_b \rightarrow \infty} \frac{n_b \sigma^2(<K>_b)}{\sigma^2(K)}
\label{eq.sigma1}
\end{equation}

where $\sigma^2(K)$ and $\sigma^2(<K>_b)$ are given by:

\begin{equation}
\sigma^2(K)=\frac{1}{M} \sum_{i=1}^M ( K_i - <K>_{total})^2
\label{eq.sigma2}
\end{equation}

\begin{equation}
\sigma^2(<K>_b)=\frac{1}{b} \sum_{j=1}^b (<K>_j - <K>_{total})^2
\label{eq.sigma3}
\end{equation}

$<K>_j$ is the average of every block and $<K>_{total}$ is the average over
the full data set of length $b*n_b=M$.

\begin{figure}[htpb]
\centerline{%
\hbox{       
   \psfig{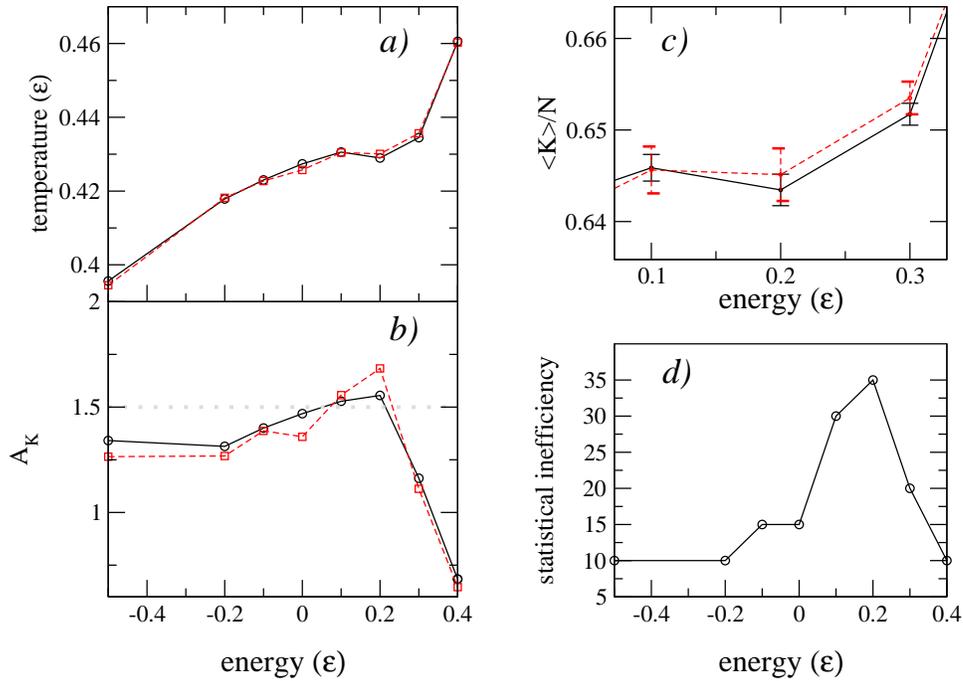}}}
\caption{\textit{In this figure we show the results of the statistical
inefficiency analysis for $\alpha=9, \rho=0.06 \sigma^{-3}$. Panel a) shows
the caloric curve for the raw data (full line, circles) and data taken via a
random sampling (dashed line, squares). Panel b) shows the relative kinetic
energy fluctuations for the same situations. In Panel c) we show more
closely the differences between the random sampling and the coarse graining
sampling (notice the errors in the random sampling case are bigger). Panel
d) shows the statistical inefficiency as a function of the energy}. }
\label{inef}
\end{figure}

Having determined the value of statistical inefficiency, the simulation runs
are divided into blocks of size $n_{b}$, and sampled either via a random
sampling, where a single value is taken at random from each block, the data
set taken as the set of such values; or coarse-graining, where the data set
is taken as the set of block averages.

In Fig.\ref{inef} we show the results of the statistical inefficiency
analysis for $\alpha =9,\rho =0.06\sigma ^{-3}$. We can see that the caloric
curve does not change appreciably, we can not state the existence of the
loop due to the numerical uncertainties involved (see panel $c$). Even
though $A_{K}$ is more sensitive to numerical correlations between the data
(it depends on both the first and second moments) this signal does not
change qualitatively between the raw data and the random sampling. It is
interesting to notice the behavior of the statistical inefficiency as a
function of the system energy: There is a clear peak in the same region
where the caloric curve displays a loop, showing that the transition from
liquid-like to a gas-like system is in correspondence with an increase in
the correlation time.

\begin{figure}[htpb]
\centerline{%
\hbox{       
   \psfig{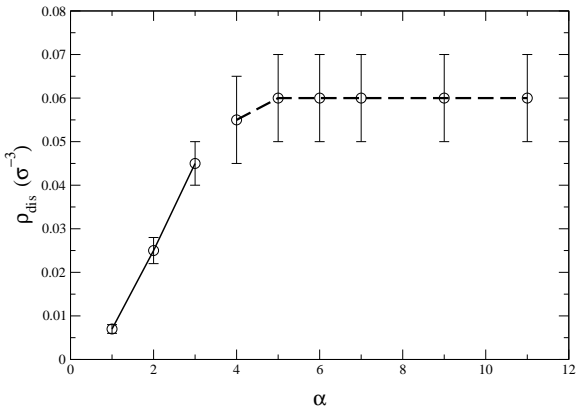}}}
\caption{\textit{Disjoint density as a function of the potential range for $%
d=3$}. }
\label{critical}
\end{figure}

Using results obtained with the above-mentioned methodology, we show in Fig.%
\ref{critical} $\rho _{dis}$ as a function of $\alpha $. We can see that
there is a clear distinction between long-range and short-range systems: For
long-range systems ($\alpha \leq 3$) $\rho _{dis}$ is an increasing function 
of $\alpha $ (a decreasing function of the potential range) which suggest that
the loop acts as a pointer of ''surfaces'', since a long-range interaction
system needs more volume to develop a structure with weakly interacting
clusters. For short-range systems there is a collapse to a $\rho
_{dis}=0.06\sigma ^{-3}$ value.

Summarizing the results of this section , we have found from our
calculations that in the case of constrained small three-dimensional systems
negative specific heats appear for all ranges of the interaction potential,
i.e. for all values of the parameter $\alpha $ for densities below an
appropriate threshold for each case.

\subsection{Two-dimensional systems}

\begin{figure}[htbp]
\centerline{%
\hbox{       
   \psfig{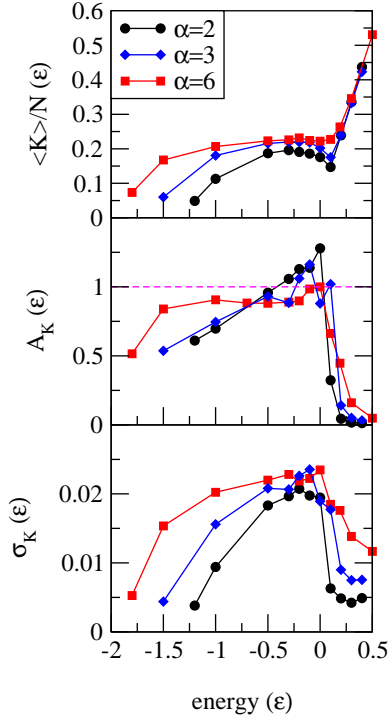}}}
\caption{\textit{Temperature, Relative fluctuations of the kinetic energy
and fluctuations of the kinetic energy as a function of the energy for $%
d=2,\rho=0.01 \sigma^{-2}$}. }
\label{kasigma}
\end{figure}

We now turn our attention to systems in two dimensions. In the upper panel of
Fig.\ref{kasigma} we show the range dependence of the caloric curves for a
constant value of the density $\rho=0.01 \sigma^{-2}$ for $N=100$ particles.

It is immediate that there are differences between the three-dimensional and
the two-dimensional case. In the former we found that if the density is low
enough, then a loop in the Caloric Curve is present for every $\alpha $ ($%
\rho _{dis}$ being an increasing function of $\alpha$). On the other hand,
 for the two-dimensional case we found a loop for a long-range potential 
($\alpha =2$), and also for a short-range case ($\alpha =3$, that correspond 
to a short-range potential in $d=2$). However the increase of $\alpha $ towards 
bigger values makes the depth of the loop almost vanish, which forbids us to 
identify negative values of $C_{v}$ due to the numerical uncertainties involved.
Moreover, If we only take into account the range of the potential $\frac{%
\alpha }{d}$ (as was defined in \cite{tsa1}) and we compare what happens for 
$two$-dimensional and $three$-dimensional systems, we realize that we have two
different behaviors for the same value of the potential range. Recall from
Fig.\ref{cc3d} that the curve corresponding to $\alpha =9$, $d=3$ exhibit a
clear loop; on the other hand from Fig.\ref{kasigma} the case $\alpha =6
$, $d=2$ presents no loop. This show that the existence of the loop is not
only related to the range of the potential, but to the dimension of the
system as well.

\begin{figure}[htbp]
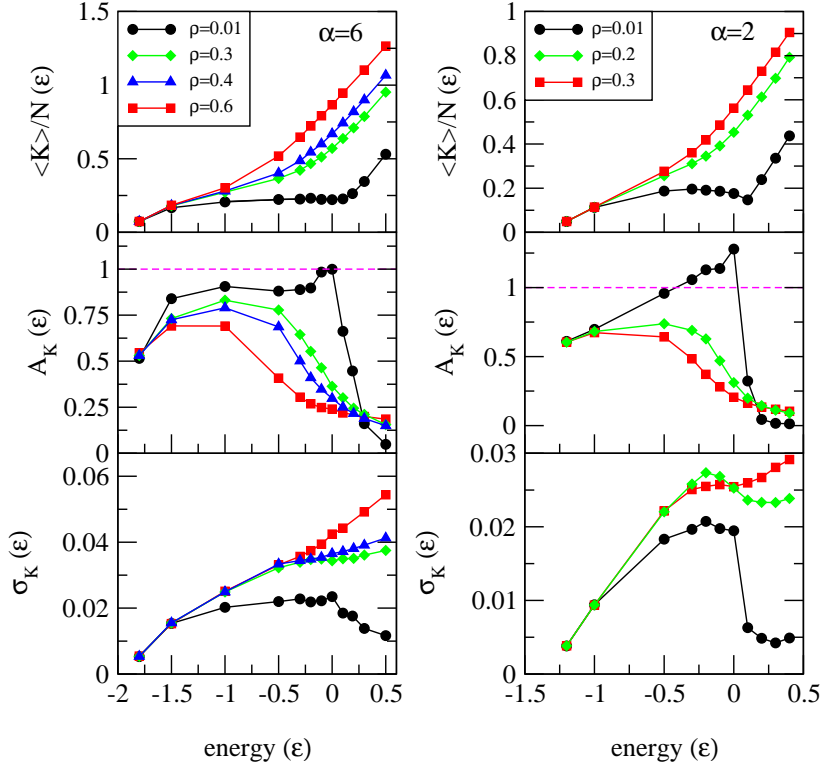

\centerline{%
\hbox{       
   \psfig{file=fig7.eps,height=4in,angle=0,clip=}
   \psfig{file=fig8.eps,height=4in,angle=0,clip=}}}
\caption{\textit{Temperature, Relative fluctuations of the kinetic energy,
potential energy and standard deviation of the kinetic energy per particle
as a function of the energy for $d=2$}.}
\label{kasigma_2d}
\end{figure}

In brief, at variance with the three-dimensional case, in two dimensions the
system does not display negative specific heats for all ranges of the
interaction potential but only for $\alpha \leq 3$, partially confirming the
results of \cite{tsa1}.

We now turn our attention to the analysis of the behavior of the quantity $%
A_{k}$ for the two-dimensional case. In the middle and lower panels of Fig.%
\ref{kasigma} we show the relative fluctuations of the kinetic energy $A_{K}$
and the standard deviation of the kinetic energy per particle as a function
of the energy for $d=2,\rho =0.01\sigma ^{-2}$. We see once again that $A_{K}
$ is a good signature of the presence of the loop. On the other hand, $%
\sigma $ is essentially the same for all curves shown (having and not having
a loop). In the $3D$ $LJ$ case, it has been found \cite{cheba} that this
signal changes from displaying a loop to a monotonous increasing function of
the energy at the critical density $\rho _{c}$. This picture is compatible
with the left panel of Fig.\ref{kasigma_2d}, where we show the density
dependence of the caloric curves, $A_{K}$, and the standard deviation of the
kinetic energy per particle for a short-range ($\alpha =6$, left panel) and
a long-range ($\alpha =2$, right panel) case. It is worth to mention that it
is only for the $\alpha =6$ system that we can make a correspondence with
the phase diagram of a $LJ$ fluid. Nevertheless the right panel of Fig.\ref
{kasigma_2d} show us that the scenario does not change qualitatively for
long-range potentials between $d=2$ and $d=3$.

\section{Conclusions}

In this work we have undertaken a detailed numerical analysis of the
thermodynamic behavior of finite confined systems in two an three dimensions
at fixed $NVE$. Our goal was to study the behavior of the specific heat in
terms of the density and dimension of the system. We identified negative
specific heats by searching for loops in the corresponding caloric curves,
and studying the relative kinetic energy fluctuations of the system, finding
consistent results in both approaches. Special emphasis was taken in the
statistical treatment of data due to correlations present in the time
series. As a consequence of the above mentioned analysis we have found that
the presence of negative specific heat is a quite general feature present in
confined systems in two an three dimensions interacting with long-ranged and
short-ranged interaction potentials. Using a LJ generalized potential in
which the range is given when the value of the parameter $\alpha $ is fixed
we have found that for three-dimensional systems there is always a value of
the density for which the system displays a negative $C_{v}$. On the other
hand, when the system is two-dimensional, there is a $\alpha _{\lim }$ such
that above it the size of the fluctuations does not allow us to identify the
presence of negative values of $C_{v}$.

\end{document}